# Observation of quantum oscillations in FIB fabricated nanowires of topological insulator (Bi$_2$Se$_3$)

**Biplab Bhattacharyya**[1,2], **Alka Sharma**[1,2], **V P S Awana**[1,2], **A. K. Srivastava**[1,2], **T. D. Senguttuvan**[1,2] and **Sudhir Husale**[1,2]*

[1]Academy of Scientific and Innovative Research (AcSIR), National Physical Laboratory, Council of Scientific and Industrial Research, Dr. K. S Krishnan Road, New Delhi-110012, India.

[2] National Physical Laboratory, Council of Scientific and Industrial Research, Dr. K. S Krishnan Road, New Delhi-110012, India.

*E-mail: husalesc@nplindia.org

# Abstract

Since last few years, research based on topological insulators (TI) is in great interests due to intrinsic exotic fundamental properties and future potential applications such as quantum computers or spintronics. The fabrication of TI nanodevices and study on their transport properties mostly focused on high quality crystalline nanowires or nanoribbons. Here we report robust approach of $Bi_2Se_3$ nanowire formation from deposited flakes using ion beam milling method. The fabricated $Bi_2Se_3$ nanowire devices have been employed to investigate the robustness of topological surface state (TSS) to gallium ion doping and any deformation in the material due to fabrication tools. We report the quantum oscillations in magnetoresistance curves under the parallel magnetic field. The resistance versus magnetic field curves have been studied and compared with Aharonov-Bohm (AB) interference effects which further demonstrate the transport through TSS. The fabrication route and observed electronic transport properties indicate clear quantum oscillations and can be exploited further in studying the exotic electronic properties associated with TI based nanodevices.

# 1. Introduction

The layered $Bi_2Se_3$ material has emerged as 3D topological insulators (TIs) [1] demonstrating exotic electronic properties due to strong spin orbit coupling and shows protected gapless metallic surface states (of massless particles, Dirac fermions) shielded from back scattering or disorders. The angle resolved photoemission spectroscopy (ARPES) has been used extensively to study the topological surface states (TSS) [2, 3]. The surface properties show features like chirality, topologically protected states due to strong spin orbit interactions, 2D

Dirac fermions with spin –momentum locking and suppressed backscattering. Though TSS of $Bi_2Se_3$ have been demonstrated through ARPES [2, 4, 5] but accessing similar properties at nanodevice level, mainly through electronic transport studies is less accessible and at the same time is complimented with more difficult complex fabrication processes. Considerable efforts have been made in order to study surface state transport at low temperature. But it is very difficult to separate residual bulk carrier conduction from surface conduction. Due to crystal defects or thermal excitations bulk carriers dominate the conduction in small bulk bandgap TI materials like $Bi_2Se_3$, thus masking the transport of 2D surface electrons [6, 7]. TIs based transport devices when shrinks down to few hundred nanometre scales may give rise to new novel phenomena due to quantum confinement effects which start dominating the transport properties. Previously quantum oscillations have been studied in TIs based materials as $Bi_2Se_3$ nanoribbons [6], $Bi_2Te_3$ nanoribbons [8], $Bi_2Te_3$ nanowires [9], SnTe nanowire [10] etc. Note that most of such studies have been performed on very high crystalline quality nanostructures of TIs and synthesized using methods like chemical vapour deposition (CVD), electrochemical deposition, vapour-liquid-solid (VLS), etc where robustness of TSS is less accessible assuming no deformation or doping in the material. Recently a study was carried out on $Bi_2Te_3$ nanotubes grown by a solution phase method where robustness of TSS against strong disorder was first time demonstrated [11]. Here we use FIB fabricated nanowires of $Bi_2Se_3$ where doping of Ga ions, deformation and cracks due to ion milling process and impurities introduced during electrode deposition by gas injection system could test the robust nature of surface states in TI materials and shows FIB fabrication technique as an efficient method for fabricating TI nanostructures. The robust nature of TSS is one of the remarkable manifestation [12] and there is an enormous interest in understanding and demonstrating it. Many fruitful studies have been done by exposing material to different doping or treatments. Cha et al., have studied the effect of magnetic doping on transport properties of $Bi_2Se_3$ nanoribbons [13]. The environmental doping, native oxide growth on $Bi_2Se_3$ crystals and nanoribbons have been shown affecting the surface degradation of the material [14]. Recently TSS of Bismuth nanoribbon was investigated for its robustness to oxidation [15] and the $Bi_2Te_3$ films were also investigated for their robust nature to silver impurities [16]. Note that TSS might get affected during the device fabrication processes such as plasma exposures, polymer contaminants, ion milling etc. and demonstration of topological properties needs more research work to be triggered. To date experimental studies showing ion beam milling and FIB nanodevice fabrication approaches are limited for investigating the TSS.

Here we report low temperature periodic quantum oscillations in FIB fabricated nanowire of $Bi_2Se_3$ under parallel magnetic field applied along the length of nanowire. The origin of these oscillations is attributed to quantum interference of time reversed paths on the surface of TI materials which encircle the nanowire in closed loops. The dominance of $h/e$ period in oscillations signifies the presence of Aharonov-Bohm interference effect in $Bi_2Se_3$ nanowire. This observation helps us to demonstrate the robust nature of TSS to FIB fabrication technique wherein gallium (Ga) ions are implanted in the sample during milling procedure. Also our fabrication method paves the way for future TIs based nanodevices with very large surface-to-volume ratio for effective surface state transport.

## 2. Methods

We have employed a very simple and straight forward technique of focused ion beam (FIB, Zeiss Auriga) milling to fabricate the nanowires of topological insulators ($Bi_2Se_3$) using micromechanically exfoliated flake as a starting material [17]. First $Si/SiO_2$ chips were cleaned with acetone, isopropanol, methanol and DI water. The oxygen plasma (Euro plasma) treatment was performed on these cleaned chips and the flakes of $Bi_2Se_3$ were deposited on the cleaned chips by using a simple micromechanical cleavage method (scotch tape). Very thin flakes were localized under an optical microscope (Olympus MX51) and later followed by a field emission scanning electron microscope (FESEM) as shown in the figure 1(a). To get the nanowire, flake was milled out by using Ga ions (figure 1(b)). The deposited $Bi_2Se_3$ flake was never exposed to ion beam (Ga) imaging. The low milling current (50pA) was used to etch some portion of the deposited flake (figure 1(b)) and the beam exposure time was less than 20s. Importantly to note that, exfoliated flake from the crystal as starting material would be of very good quality and controlled FIB milling can fabricate nanostructures instantly with desired shapes and sizes. Metal electrodes of platinum (Pt) were deposited immediately on the fabricated nanowire with the help of precursor based gas injection system (GIS). Recently we have studied physical parameters such as carrier concentration, mobility etc. of FIB fabricated nanowires of topological insulators ($Bi_2Se_3$) material [18]. High resolution transmission electron microscopy (HRTEM model: Tecnai G2 F30 STWIN with field emission gun and an electron accelerating voltage of 300 kV) was used to characterize $Bi_2Se_3$ samples for microstructural information even up to atomic scale including the reciprocal space (figure 2). Overall, a thin flake-type morphology having a facetted-edge was observed throughout in the microstructure. Figure 2(a) & (b) display several flakes of nano-dimensions. The thicknesses of these flakes are expected between 20 to 50 nm. Moreover the stacking of the nano-flakes over each other is quite visible in the microstructure. The dotted encircled regions reveal the stacking of at least three layers of the flakes with sharp-long edges. Moreover the flakes are also noted physically assembled over each other. The contrast evolved within the individual flakes could be attributed to several reasons, like the strain developed during nucleation and growth, grain boundaries, mechanical strain, etc. However a set of white arrows (figure 2(b)) reveals a contrast evolved due to the edge sharing of nanocrystallites in these flaky-microstructures. This evidence has become obvious at high magnifications, where a set of moiré fringes with the fringe spacing of about 0.63 nm are displayed (inset figure 2(b)). A set of atomic scale images recorded from different regions of the flaky-microstructure shows that the flakes are crystalline in nature. A stacking of $10\bar{1}1$ planes of a rhombohedral phase of $Bi_2Se_3$ (space group: $R\bar{3}m$, lattice constants: a = b = 0.414 nm, c = 2.863 nm, ref.: JCPDS card no. 33-0214) with the interplanar spacing of 0.36 nm, exhibits good crystallinity in real space (figure 2(c)). Similarly inset in figure 2(c), shows a set of cross-over atomic planes of $Bi_2Se_3$ along $10\bar{1}1$ and $01\bar{1}5$ with the corresponding interplanar spacings of 0.36 and 0.30 nm, respectively. In the reciprocal space, a selected area electron diffraction pattern (SAEDP) from an assembly of nano-flakes displays a ring pattern consisting of Debye rings evolved due to the different atomic planes of $Bi_2Se_3$. Some of these important planes: $10\bar{1}1$, $01\bar{1}5$, $10\bar{1}10$ and $11\bar{2}0$ $Bi_2Se_3$ rhombohedral crystal structure with the interplanar spacings of 0.36, 0.30, 0.23, and 0.21, respectively, are marked on the SAEDP (figure 2(d)). In addition, inset in figure 2(d) exhibits another SAEDP recorded along [$01\bar{1}2$] zone axis with the denotation of $01\bar{1}\bar{1}$ and $2\bar{1}\bar{1}0$ crystallographic planes of rhombohedral structure.

## 3. Results and Discussion

The fabricated devices (S2, N15 and N4) having different widths (185nm, 334nm and 140nm) and lengths (0.5μm, 3.96μm and 2.4μm) of the nanowire were studied at low temperature using a physical properties measurement system (PPMS by Quantum design). Figure 3(a I) shows the schematic diagram of $Bi_2Se_3$ nanowire under parallel magnetic field (B) and current (I) applied along the length of nanowire. The red cone represents the Dirac nature of surface states and two green arrows in same loop but in opposite direction represents the two partial electron waves that encircle the same magnetic flux and interfere on the other side of nanowire. Surface electron transport states showing Aharonov-Bohm interference effects (h/e oscillations) have been studied previously in TIs based nanowires or nanofilms when the magnetic field was applied parallel to the flow of electrons in the device [6, 8, 9, 19, 20]. Here fabricated nanowires based devices are tested for an external B field applied along the flow of current (length of the nanowire). It is expected that interference effects can be observed only when the conduction of electron remains phase coherent while completing the closed trajectories and periodic MR oscillations with $h/e$ flux quantization is a hallmark of the AB interference effect [6]. The period of oscillation is related to B field as $\Delta B = \phi_0/S$, where S is the cross sectional area of the nanowire and $\phi_0$ is the flux quantum ($h/e$). In our case due to ion milling process the nanowire gets deformed and Ga ions are implanted in the material as shown in a schematic of nanowire cross section (figure 3(a II)). We propose that the presence of strong spin-orbit coupling in TI materials does not allow the electron waves to travel through the highly disordered/deformed trajectories as it may disturb the time reversal symmetry and enhance backscattering. Thus electrons encircle the nanowire circumference in a completely different trajectory where disorder is less (shown by a green path in figure 3(a II)). This decreases the effective cross sectional area enclosed by electron waves which leads to an increase in flux quantisation period in our devices. To observe such effects we have performed MR measurements on device S2 (figure 3(c)), N15 (figure 4(b)) and N4 (figure 5). The observed nonlinear MR in these samples in the presence of parallel B field clearly supports the property of investigating TSS. The device S2 exhibits more semiconducting like behaviour at low temperature (figure 3(b)). For this device, RT curve shows gradual decrease in resistance on cooling from room temperature down to ~ 35 K. The increase in resistance is observed for the temperature range of 28 to 2 K. Inset (figure 3(b)) shows the logarithmic plot of the resistance with the inverse of temperature (1/T). The plot is used to get the estimation of activation energy for the temperature range 28K to 2K. Here we used the relation $R \sim e^{E_a/k_B T}$ where $R$ is the resistance, $E_a$ is activation energy, $k_B$ is the Boltzmann constant and T is temperature. From the slope of the curve we estimate the activation energy ~2.2 meV. Earlier RT measurements done on nanoribbons of $Bi_2Te_3$ also showed small activation energy which is consistent with our observed transport properties [8]. The oscillations in resistance for sample S2 were studied at different constant temperatures (2K, 6K, 10K, 15K and 20K) which are shown in figure 3(c). The non linear MR, periodic oscillations and decrease in amplitude with increase in temperatures are clearly evident as shown in figure 3(c). The black curve (2K) represents clear and reproducible oscillations in R with periodicity of ΔB about 0.9T. The slight variation in the perfect periodicity has been noticed but on an average 0.9T is observed between two minima in resistance as highlighted by the dotted line on 2K curve. For S2, taking into account of AB effect, this period corresponds to a cross sectional area of about 4580 nm$^2$. The estimated cross section area of device S2 is about 5400 nm$^2$ which is slightly deviating (assuming 15% error arising from surface morphology in experimental measurements of the area of nanodevices). Also it could be the sample response due to FIB milling which might slightly deform the material or any doping of gallium atom impurities. The slight deterioration in perfect periodicity has been also observed previously at higher fields [6] and it could be due to contributions from the Shubnikov-de Haas (SdH) oscillations of the bulk or aperiodic universal conductance fluctuations (UCFs) [21] arising from bulk states superimposing on the AB oscillations. Additionally theoretical work by Bardarson et al. pointed out that conductance oscillations with magnetic flux of $h/e$ or $h/2e$ depend on doping and disorder strength and indicated the motion of electrons in diffusive regime [22]. The surface states have been further probed by plotting the temperature (T) dependence of Fast Fourier transform (FFT) amplitude of dR/dB in sample S2 (figure 3(d)). For the temperature range 2K to 20K inelastic phonon scattering is absent as indicated by a $T^{-1/2}$ fit of the FFT amplitudes [8]. The phase coherence length $l_\varphi = \sqrt{D\tau_\varphi}$ where $\tau_\varphi \sim \hbar/k_B T$ is the average time between inelastic events and D is the diffusion constant. Here we observed a dominant peak which corresponds to $h/e$ oscillations (inset figure 3(d)) but the second harmonic peak $h/2e$ was found difficult to distinguish from the other peaks. Figure 4 depicts the RT and MR measurements performed for device N15 which shows a metallic behaviour while cooling down till 2K (figure 4(a)). The periodic oscillations in MR at 6K were noticed with periodicity of ~2.4T between two minima or maxima observed for this sample N15. The actual cross sectional area for device N15 is ~9700 nm$^2$. The observed period corresponds to a cross section area of 1725nm$^2$ of nanowire (80% deviation from expected result). We see an increase in period with increase in width (or area) of the nanowire. This result is unexpected as period should decrease with increase in width of nanowire. The quadratic like increase in the resistance was noticed at low field as shown in the inset (figure 4(b)) and arrow represents small cusp like feature formation at zero field, even though it is not sharp compared to earlier reports but could be the signature of weak anti-localization (WAL) or Dirac fermions [6].

In order to test the robustness of TSS in FIB fabricated Bi$_2$Se$_3$ devices we performed MR measurements for one more sample (N4) having smaller width ~ 140 nm compared to other two devices studied. Figure 5 shows the periodic quantum oscillations in MR with period of ~2.5T at 15K for this device. Interestingly to note that oscillations were clearly visible even at temperature 15K which is could be the robust nature and an enhancement oscillation due to increase in surface to volume ratio. The MR minima was observed on integral multiples of $h/e$ (dashed orange line in figure 5) which clearly suggests that these quantum oscillations are a result of 0-AB (oscillations at zero flux have a minimum) effect due to metallic surface states in Bi$_2$Se$_3.$ The FFT amplitude spectrum (inset figure 5) further verifies the $h/e$ period of about 2.6T. A prominent $h/2e$ period (~1.25T) is visible for device N4 which was absent in FFT spectrum of device S2. The primary reason for this appearance of $h/2e$ period peak can be attributed to the small dips which superimpose on the broad peaks corresponding to 0-AB effect. These small dips occur at odd multiples of $h/2e$ (dashed grey lines in figure 5). Previous studies have shown that such a phenomenon is possible due to weak π-AB (oscillations at zero flux have a maximum) oscillations where a MR minimum appears at odd multiples of $h/2e$, indicating the π-Berry's phase in TSS [9, 23]. Altshuler-Aronov-Spivak (AAS) effect with period $h/2e$, where minima at even multiples of $h/2e$ coincide with minima of 0-AB oscillations can also give rise to such small dips in MR [9]. Note that when compare the data with AB effect we find an error of 58% between the actual cross sectional area (4000nm$^2$) and the one calculated from the observed period (1656nm$^2$) for this device. But the increased flux quantization period in N4 when compared to device S2 is a consequence of decreased width in N4 which is consistent with the formula $\Phi = B.S$ and previous studies [6]. Note that the AB and AAS effects can be observed in non-topological insulator material also if the electron transport in investigated system maintains the quantum coherency which gives rise to conductance oscillations. The earlier work on Bismuth nanowires [24], multiwall carbon nanotube [25], Au nanorings [26], magnesium film evaporated onto a quartz filament [27], graphene ring [28] and Dirac semimetal nanowires [29] have shown quantum interference of surface states indicating the AB oscillations. We assume these periodic oscillations could be coming from AB effects and gallium doping and or material deformation also affecting oscillation period and phase coherence length $l_\varphi$ of the electrons. Normally, $l_\varphi$ in Bi$_2$Se$_3$ is reported about 500nm [6, 8], but in our case it is assumed to be much shorter than nanowire circumference. Important to note that bulk carriers dominated transport expected to show fluctuations or oscillations in resistance with non-periodic dependence on the applied B field [6]. Because quantum interference effects due to random electron paths in bulk cancel out leading to aperiodic MR oscillations whereas in case of TSS interference effects do not average out as depicted by periodicity of MR oscillations. The dominance of $h/e$ period in our results demonstrates the TSS and its robustness to disorder induced by ion milling. However, TI materials or devices are prone to surface degradation affecting the mobility and very few reports demonstrated clear quantum oscillations so far, hence, more careful studies need to be carried out [30].

Previous theoretical studies have shown that amount of disorder in material affects the WAL feature and period of flux quantization by making the electron motion highly diffusive [22]. Though Ga impurities are non-magnetic in nature and cannot affect the spin-orbit coupling in TIs but can alter the phase of electron wave by introducing diffusive incoherent motion and changing the Fermi momentum ($p_F$) which affects the phase as $\theta(B) = \frac{p_F L}{\hbar} + \frac{e}{c\hbar}\int \vec{A}.\vec{dr}$ where L = path length for electron, ℏ is the reduced Planck's constant and second term is the relativistic AB phase acquired by electron in parallel magnetic field where $\vec{A}$ is the magnetic vector potential, e is electronic charge, c is the speed of light and $dr$ is the differential length element over which integration is performed [31]. Also, previous theory suggests that diffusive paths add random phases to electron wave which make the conductance oscillations sample specific. AB dephasing length $l_\varphi^{AB}$, the length over which electron phase is maintained before getting disturbed due to inelastic interactions [32], was shown proportional to $\frac{D}{\sqrt{TR}}$ with dephasing rate $\frac{1}{\tau_\varphi^{AB}} \sim \frac{TR}{D}$ where $\tau_\varphi^{AB}$=dephasing time, D=diffusion constant, T=temperature, R=radius of nanowire. Thus, increasing the nanowire width decreases the $l_\varphi$ and short phase coherence length may account for the absence of AAS effect ($h/2e$ periods) in device S2 and anomalous oscillation period in device N15. The surface potential induced by the implanted Ga ions and other impurities during FIB milling or metal deposition process may play a significant role in tuning the position of Dirac point and electron distribution near the surface and at the interface between TI and implanted material. Recently Men'shov V N et al. [33] have shown the detailed analytical and numerical explanation for overlayer induced surface potential in a 3D TI/normal insulator type heterostructure where the surface potential matrix elements ($U_i$) can be related to the work functions ($\Phi$), bandgap energies (E) of the overlayer and TI as $U_{overlayer} \sim \Phi_{TI} - \Phi_{overlayer} + E_{overlayer} - E_{TI}$ and $U_{TI} \sim \Phi_{TI} - \Phi_{overlayer}$. The strength and sign of surface potential depend on the energy dispersion and spatial distribution of the electrons near the surface. It is expected that for weak surface potential, Dirac point shifts slightly up or down depending on the sign, whereas for strong surface potential Dirac point is pushed deep in the bulk valence band. In another report Eremeev et al. [34], studied the surface band structure of layered 3D TI material ($Bi_2Se_3$, $Bi_2Te_3$) as a function of adsorbate deposition time and impurity atom size. They proposed an expansion of the outermost van der Waals (vdW) gap in between two consecutive quintuple layers due to impurity intercalation which leads not only to the emergence of Rashba-split parabolic 2D electron gas (2DEG) bands below the bulk conduction band and M-shaped 2DEG bands in the bulk valence band, but also the relocation of TSS to the lower quintuple layer. Further, it was observed that the TSS relocation process does not affect the dispersion of upper part of Dirac cone and position of the Dirac point but the lower part shows some variations. This helps in understanding our observed experimental proposition that electrons travel in a completely different trajectory where deformation and contamination due to Ga milling and GIS is less which accounts for the decrease in effective area of cross section and increase in flux quantization period. The observation of weak localization (WL) in TIs was previously reported for ultrathin films in which top and bottom surface wave functions interfere to open a gap in surface states [35]. In $Ar^+$ ion implementation study it was observed that the implementation of $Ar^+$ ions increases the disorder by increasing defect states in TI band structure. Localized states are created due to disorder and electron transport occurs due to hopping across these states. The competition between conventional electron transport in conduction band of TIs and electron transport due to hopping results in WL [36].

## 4. Conclusion

In conclusion, the experimental analysis of MR demonstrates periodic oscillations in resistance indicating the existence of TSS in FIB fabricated $Bi_2Se_3$ nanoscale devices. Generally in TIs, WAL effect is observed around zero magnetic field which is the destructive interference of two electron waves [37, 38] and similar was also observed in sample N15 (small cusp at 0T in figure 4(b)). A primary $h/e$ period peak in FFT spectrum (inset figure 3(d)) for S2 shows the dominance of AB effect in resistance fluctuations under parallel magnetic field. The linear MR was totally absent for all the samples under parallel field which indicate the surface state effect. The disappearance of periodic oscillations with increasing temperature (figure 3(c)) indicates their quantum nature and also the increased interference of bulk states with surface conduction at high temperature. The oscillations become damped and aperiodic due to high bulk electron-electron interaction and electron-phonon interaction [6, 8, 32]. Overall, we consider these periodic quantum oscillations to be a remarkable manifestation of robust nature of TSS. Moreover, the VLS method is more common to synthesize the nanostructures of topological insulators [39]. But compared to it alternatively, FIB fabrication has some advantages such as pattering the high quality flake to desired shape, size, dot, ring or even 3D structure. Further fabricated nanostructures can be in-situ transform into a nanodevice used in the quantum transport measurement if the electrical metal contacts are made by using the precursor based gas injection system like (Pt/Au), superconductor (W) or ferromagnet (Co) which gives more opportunities for investigating the novel quantum transport phenomena predicated with TIs materials. Here we believe that nanowire fabrication by FIB method provides a very efficient way to check for the robustness of TSS against doping induced disorders.

## Acknowledgements

S.H. and A.S. acknowledge the support of CSIR's Network project "Aquarius". B.B. acknowledges the support of JRF fellowship from CSIR. We thank Dr. Sangeeta Sahoo for the critical reading of the manuscript.

## References

1. H. Zhang, C.-X. Liu, X.-L. Qi, X. Dai, Z. Fang and S.-C. Zhang, Nat Phys **5** (6), 438-442 (2009).
2. Y. Xia, D. Qian, D. Hsieh, L. Wray, A. Pal, H. Lin, A. Bansil, D. Grauer, Y. S. Hor, R. J. Cava and M. Z. Hasan, Nat Phys **5** (6), 398-402 (2009).
3. D. Hsieh, Y. Xia, D. Qian, L. Wray, J. H. Dil, F. Meier, J. Osterwalder, L. Patthey, J. G. Checkelsky, N. P. Ong, A. V. Fedorov, H. Lin, A. Bansil, D. Grauer, Y. S. Hor, R. J. Cava and M. Z. Hasan, Nature **460** (7259), 1101-1105 (2009).
4. S.-Y. Xu, Y. Xia, L. A. Wray, S. Jia, F. Meier, J. H. Dil, J. Osterwalder, B. Slomski, A. Bansil, H. Lin, R. J. Cava and M. Z. Hasan, Science **332** (6029), 560 (2011).
5. C. Chen, S. He, H. Weng, W. Zhang, L. Zhao, H. Liu, X. Jia, D. Mou, S. Liu, J. He, Y. Peng, Y. Feng, Z. Xie, G. Liu, X. Dong, J. Zhang, X. Wang, Q. Peng, Z. Wang, S. Zhang, F. Yang, C. Chen, Z. Xu, X. Dai, Z. Fang and X. J. Zhou, Proceedings of the National Academy of Sciences of the United States of America **109** (10), 3694-3698 (2012).
6. H. Peng, K. Lai, D. Kong, S. Meister, Y. Chen, X.-L. Qi, S.-C. Zhang, Z.-X. Shen and Y. Cui, Nat Mater **9** (3), 225-229 (2010).
7. B. Hamdou, J. Gooth, A. Dorn, E. Pippel and K. Nielsch, Applied Physics Letters **102** (22), 223110 (2013).
8. F. Xiu, L. He, Y. Wang, L. Cheng, L.-T. Chang, M. Lang, G. Huang, X. Kou, Y. Zhou, X. Jiang, Z. Chen, J. Zou, A. Shailos and K. L. Wang, Nat Nano **6** (4), 216-221 (2011).
9. M. Tian, W. Ning, Z. Qu, H. Du, J. Wang and Y. Zhang, Sci. Rep. **3** (2013).
10. M. Safdar, Q. Wang, M. Mirza, Z. Wang, K. Xu and J. He, Nano Letters **13** (11), 5344-5349 (2013).
11. R. Du, H.-C. Hsu, A. C. Balram, Y. Yin, S. Dong, W. Dai, W. Zhao, D. Kim, S.-Y. Yu, J. Wang, X. Li, S. E. Mohney, S. Tadigadapa, N. Samarth, M. H. W. Chan, J. K. Jain, C.-X. Liu and Q. Li, Physical Review B **93** (19), 195402 (2016).
12. J. Lee, J.-H. Lee, J. Park, J. S. Kim and H.-J. Lee, Physical Review X **4** (1), 011039 (2014).
13. J. J. Cha, J. R. Williams, D. Kong, S. Meister, H. Peng, A. J. Bestwick, P. Gallagher, D. Goldhaber-Gordon and Y. Cui, Nano Letters **10** (3), 1076-1081 (2010).

14. D. Kong, J. J. Cha, K. Lai, H. Peng, J. G. Analytis, S. Meister, Y. Chen, H.-J. Zhang, I. R. Fisher, Z.-X. Shen and Y. Cui, ACS Nano **5** (6), 4698-4703 (2011).
15. W. Ning, F. Kong, Y. Han, H. Du, J. Yang, M. Tian and Y. Zhang, Sci. Rep. **4** (2014).
16. T. Zhang, P. Cheng, X. Chen, J.-F. Jia, X. Ma, K. He, L. Wang, H. Zhang, X. Dai, Z. Fang, X. Xie and Q.-K. Xue, Physical Review Letters **103** (26), 266803 (2009).
17. A. Sharma, B. Bhattacharyya, A. K. Srivastava, T. D. Senguttuvan and S. Husale, Scientific Reports **6**, 19138 (2016).
18. B. Bhattacharyya, A. Sharma, V. P. S. Awana, T. D. Senguttuvan and S. Husale, arXiv:1611.00564 (2016).
19. L. A. Jauregui, M. T. Pettes, L. P. Rokhinson, L. Shi and Y. P. Chen, Sci. Rep. **5** (2015).
20. S. S. Hong, Y. Zhang, J. J. Cha, X.-L. Qi and Y. Cui, Nano Letters **14** (5), 2815-2821 (2014).
21. P. A. Lee and A. D. Stone, Physical Review Letters **55** (15), 1622-1625 (1985).
22. J. H. Bardarson, P. W. Brouwer and J. E. Moore, Physical Review Letters **105** (15), 156803 (2010).
23. Y. Zhang and A. Vishwanath, Physical Review Letters **105** (20), 206601 (2010).
24. A. Nikolaeva, D. Gitsu, L. Konopko, M. J. Graf and T. E. Huber, Physical Review B **77** (7), 075332 (2008).
25. A. Bachtold, C. Strunk, J.-P. Salvetat, J.-M. Bonard, L. Forro, T. Nussbaumer and C. Schonenberger, Nature **397** (6721), 673-675 (1999).
26. R. A. Webb, S. Washburn, C. P. Umbach and R. B. Laibowitz, Physical Review Letters **54** (25), 2696-2699 (1985).
27. D. Y. Sharvin and Y. V. Sharvin, JETP Lett. **34** (5), 272-275 (1981).
28. S. Russo, J. B. Oostinga, D. Wehenkel, H. B. Heersche, S. S. Sobhani, L. M. K. Vandersypen and A. F. Morpurgo, Physical Review B **77** (8), 085413 (2008).
29. L.-X. Wang, C.-Z. Li, D.-P. Yu and Z.-M. Liao, Nature Communications **7**, 10769 (2016).
30. J. J. Cha, K. J. Koski and Y. Cui, Physica Status Solidi-Rapid Research Letters **7** (1-2), 15-25 (2013).
31. I. L. Aleiner, A. V. Andreev and V. Vinokur, Physical Review Letters **114** (7), 076802 (2015).
32. T. Ludwig and A. D. Mirlin, Physical Review B **69** (19), 193306 (2004).
33. V. N. Men'shov, V. V. Tugushev, T. V. Menshchikova, S. V. Eremeev, P. M. Echenique and E. V. Chulkov, Journal of Physics: Condensed Matter **26** (48), 485003 (2014).
34. S. V. Eremeev, M. G. Vergniory, T. V. Menshchikova, A. A. Shaposhnikov and E. V. Chulkov, New Journal of Physics **14** (11), 113030 (2012).
35. L. Zhang, M. Dolev, Q. I. Yang, R. H. Hammond, B. Zhou, A. Palevski, Y. Chen and A. Kapitulnik, Physical Review B **88** (12), 121103 (2013).
36. K. Banerjee, J. Son, P. Deorani, P. Ren, L. Wang and H. Yang, Physical Review B **90** (23), 235427 (2014).
37. H. B. Jens and E. M. Joel, Reports on Progress in Physics **76** (5), 056501 (2013).
38. O. Pavlosiuk, D. Kaczorowski and P. Wiśniewski, Scientific Reports **5**, 9158 (2015).
39. D. Kong, J. C. Randel, H. Peng, J. J. Cha, S. Meister, K. Lai, Y. Chen, Z.-X. Shen, H. C. Manoharan and Y. Cui, Nano Letters **10** (1), 329-333 (2010).

**Figures**

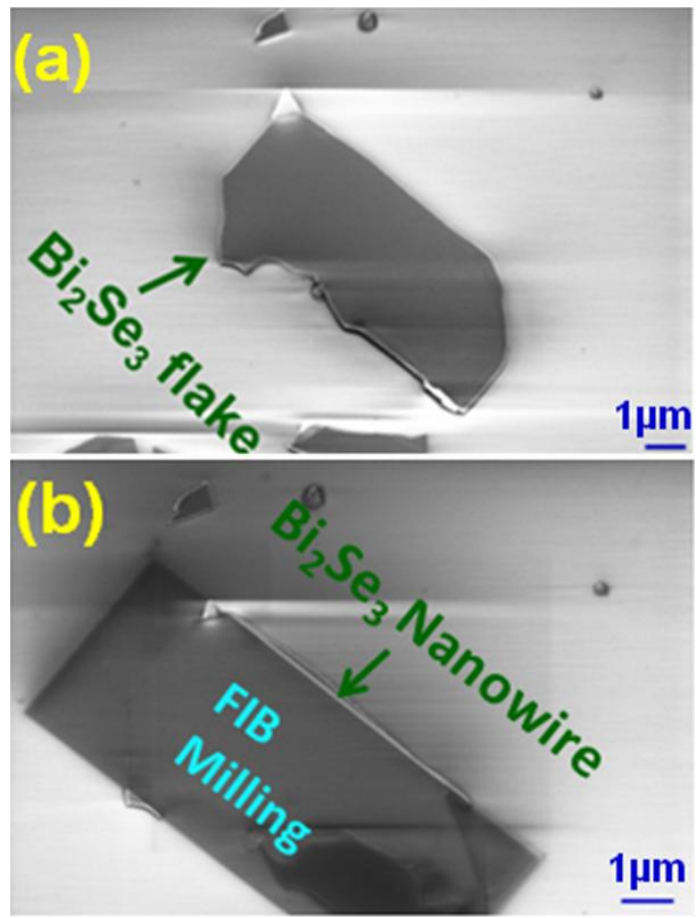

**Figure 1. Bi$_2$Se$_3$ nanowire synthesis.** FESEM image of (a) Bi$_2$Se$_3$ flake, and (b) nanowire after FIB milling. Scale bar is 1μm.

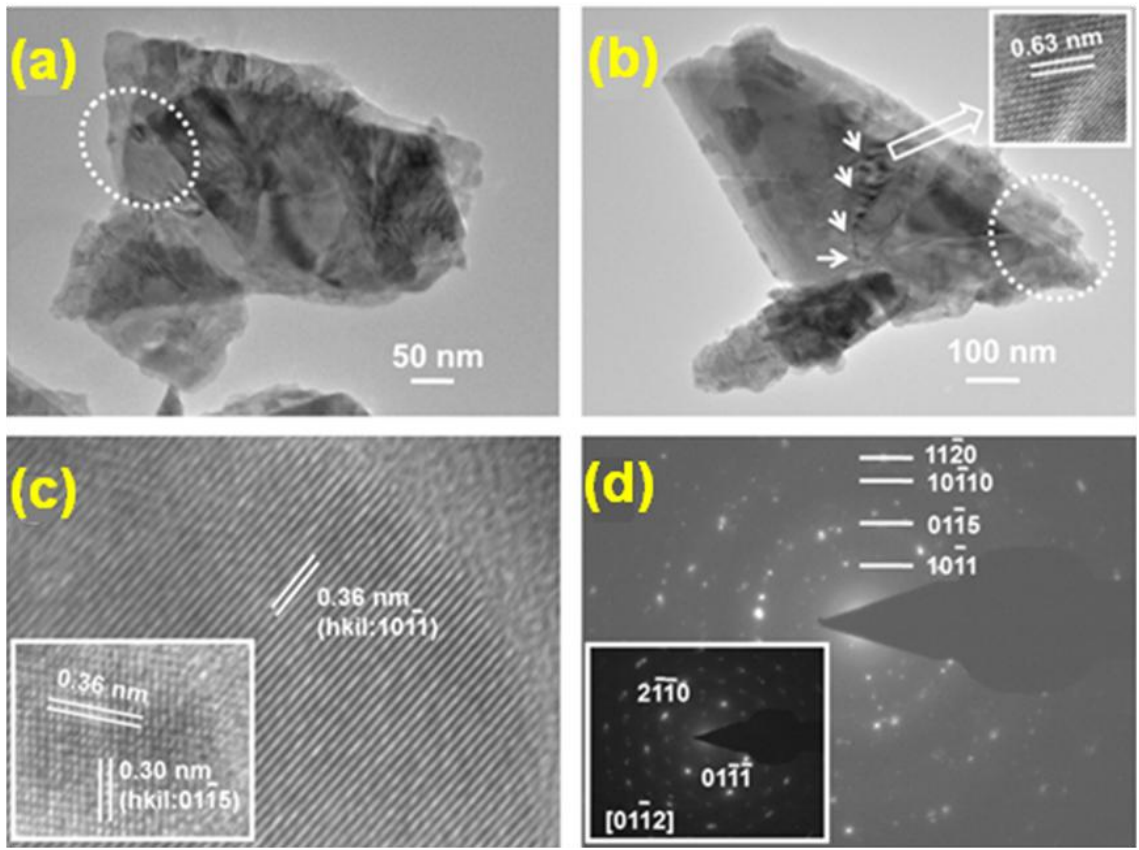

**Figure 2. HRTEM characterization.** (a-d) Microstructural and crystallographic investigations of $Bi_2Se_3$ showing, (a & b) nano-flake morphology. (c) Atomic scale image of nano-flakes. (d) SAEDP of rhombohedral crystal structure. Insets: (b) Moiré fringes, (c) atomic scale image, and (d) SAEDP of single flake morphology.

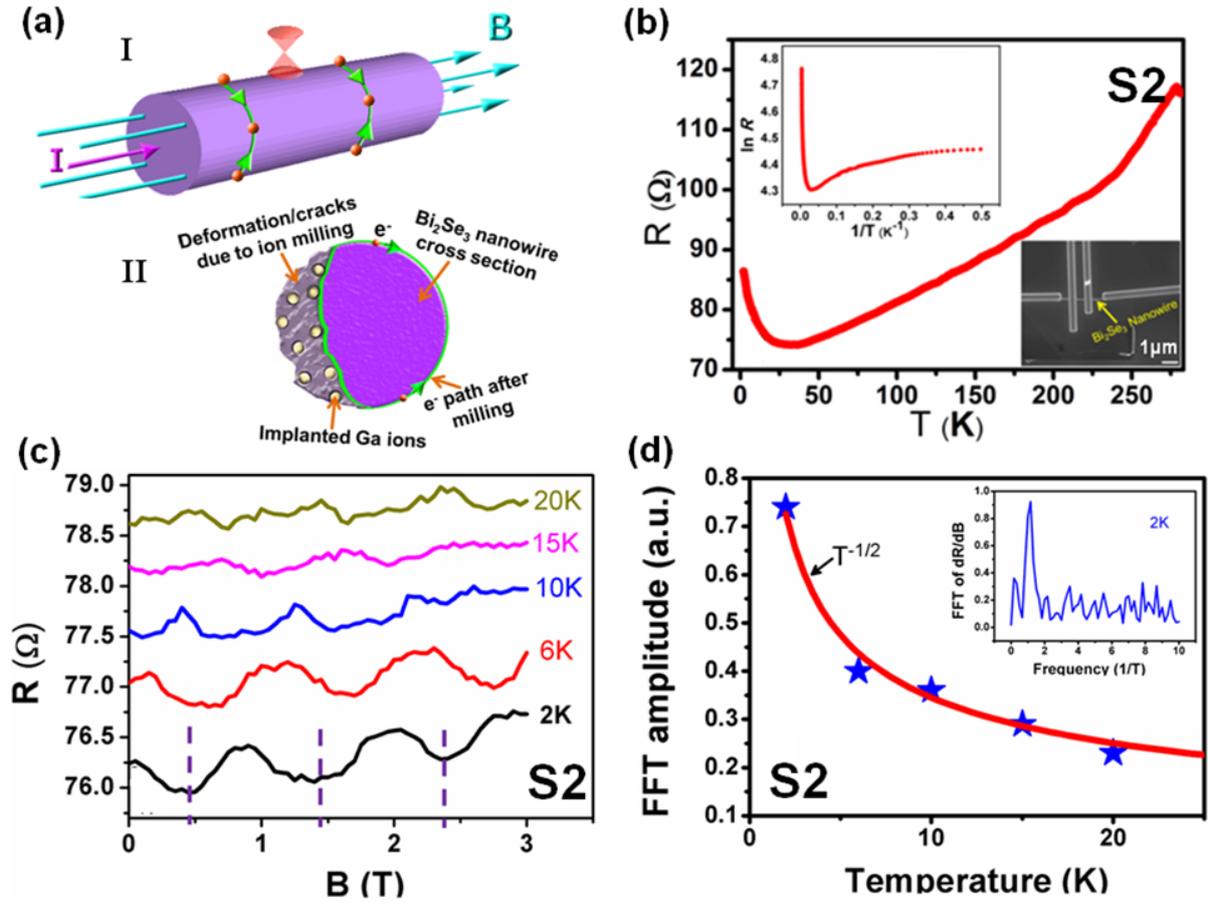

**Figure 3. Signatures of Aharonov-Bohm interference in presence of parallel magnetic field for device S2.** (a I) 3D schematic of AB interference effect in $Bi_2Se_3$ nanowire. Dirac cone is shown in red. (a II) 2D schematic of nanowire cross section shows the deformation/cracks and Ga ions introduced in the nanowire due to FIB milling. Two partial electron ($e^-$) waves encircle the magnetic flux in a different trajectory compared to conventional path as shown in green. (b) Metallic and semiconducting behaviour of RT curve, inset upper left is the logarithmic scale R versus the inverse of temp (1/T). FESEM image of the device S2 (inset down right) scale bar is 1μm and B=0. (c) MR curves measured at 2K, 6K, 10K, 15K and 20 K. The dashed lines on the black curve indicate a clear modulation of the resistance with period ~ 0.95T (B sweep 0 to 3T). (d) AB oscillations and its temperature dependence roughly show the $T^{-1/2}$ relation (red curve). FFT amplitude is measured from the observed h/e peaks. Inset shows the FFT of the dR/dB (derivative) versus inverse of the magnetic field (1/B) at 2K and the peak shows the flux quantization.

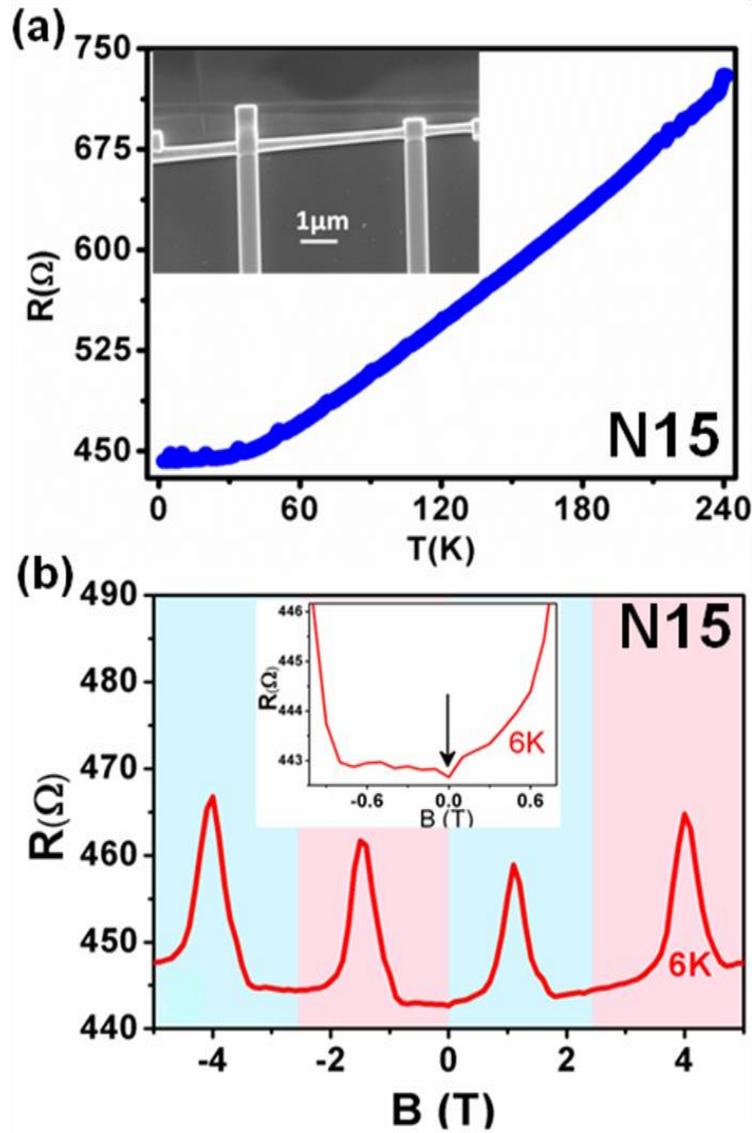

**Figure 4. RT and MR for device N15.** (a) Cooling curve shows a metallic behaviour. Inset is the FESEM image of the device. Scale bar is 1μm and B=0. (b) MR curve measured at 6K. The colour bar shades represent the two resistance minima equivalent to an oscillation period of ~ 2.5 T in magnetic field. Inset shows the magnified part of MR data at low B field. The arrow indicates the formation of a very small sharp cusp, indicating WAL features.

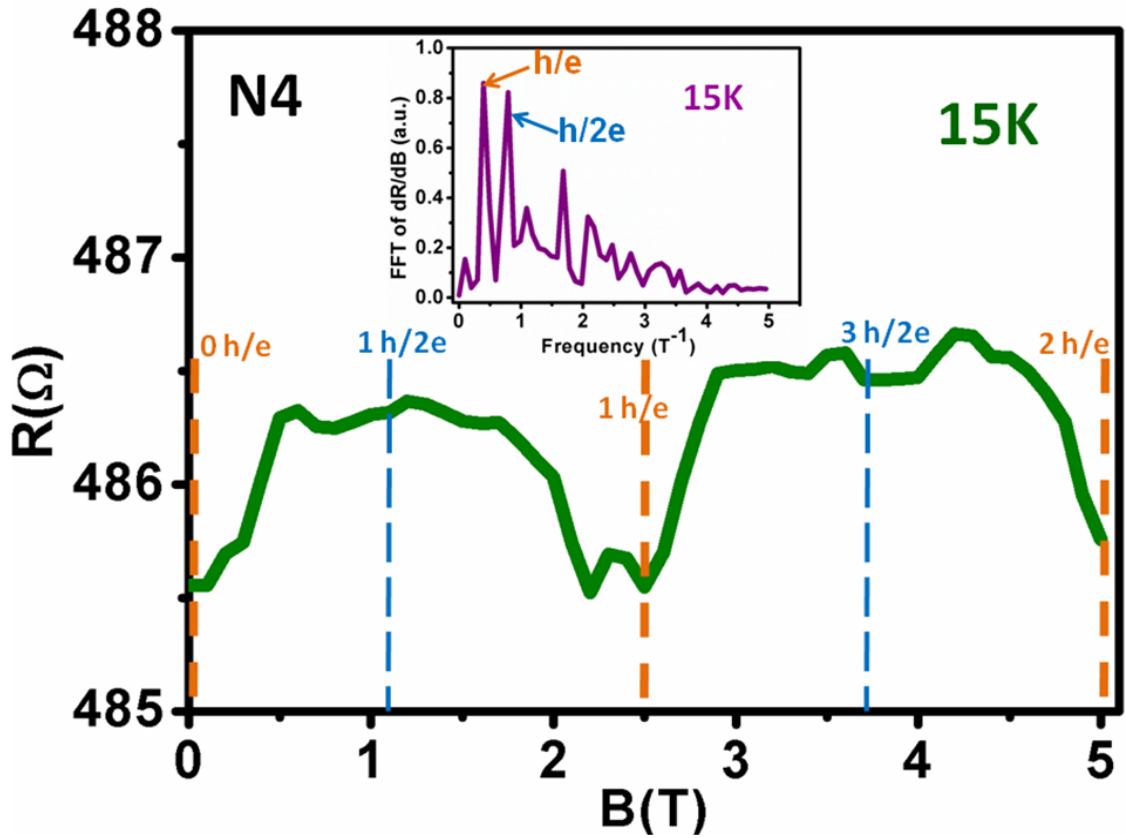

**Figure 5. Quantum oscillations in device N4.** MR curve shows clear quantum oscillations with period ~2.5T even at 15K. The minima indicated by orange dashed lines are located at integral multiples of $h/e$. Small dips which occur at odd multiples of $h/2e$ (shown in blue dashed lines) superimpose on the broad peaks. $h/e$ and $h/2e$ peaks are labelled in FFT spectrum (inset).